\documentstyle[prb,twocolumn,aps]{revtex}

\input{epsf}

\begin{document}

\tolerance 10000

\twocolumn[\hsize\textwidth\columnwidth\hsize\csname %
@twocolumnfalse\endcsname

\draft

\title{Quantum Phase Transitions and the Breakdown of Classical General
       Relativity}

\author{G. Chapline}

\address{Los Alamos National Laboratory, Los Alamos, New Mexico 87545 }

\author{E. Hohlfeld, R. B. Laughlin, and D. I. Santiago }

\address{Department of Physics, Stanford University,
        Stanford, California 94305}

\date{\today}
\maketitle
\widetext

\begin{abstract}
\begin{center}

\parbox{14cm}{It is proposed that the event horizon of a black hole is a
       quantum phase transition of the vacuum of space-time analogous
       to the liquid-vapor critical point of a bose fluid.  The
       equations of classical general relativity remain valid arbitrarily
       close to the horizon yet fail there through the divergence of
       a characteristic coherence length $\xi$.  The integrity of global
       time, required for conventional quantum mechanics to be defined,
       is maintained. The metric inside the event horizon is different
       from that predicted by classical general relativity and may be de
       Sitter space.  The deviations from classical behavior lead to
       distinct spectroscopic and bolometric signatures that can, in
       principle, be observed at large distances from the black hole.}

\end{center}
\end{abstract}

\pacs{
\hspace{1.9cm}
PACS numbers: {04.70.Dy,05.70.Jk,05.30.Jp,64.60.Ht}
}
]

\narrowtext

\section{Introduction}

Quantum mechanics is incompatible with classical general relativity. While
there are many ways of articulating the problem, all reduce in the end to
the absence of universal time required for the many-body Schr\"{o}dinger
equation

\begin{equation}
i \hbar \frac{\partial \Psi}{\partial t} = {\cal H} \; \Psi
\end{equation}

\noindent
to make sense.  This equation is the logical underpinning of quantum field
theory and statistical mechanics, and thus of our microscopic
understanding of the entire natural world outside gravity.  General
relativity predicts that certain stars evolve at the end of their lives
into black holes characterized by surfaces at which time, as measured
by a clock at infinity, stands still.  Gravity is well-behaved at this
surface, in that a free-falling observer passes through in finite proper
time without being ripped apart by tidal forces, but quantum mechanics is
not. The paradox is fundamental. It has led to proposals for revising the
laws of quantum mechanics \cite{thooft} and to speculations that black
holes may destroy quantum information \cite{info}.

In this paper we propose a resolution of this problem that is fully
quantum-mechanical and is based on principles that can be tested in the
laboratory. The essence of the idea, illustrated in Fig. 1, is that the
black hole event horizon is a continuous quantum phase transition of the
vacuum of space-time roughly analogous to the quantum liquid-vapor
critical point of an interacting bose fluid.  In such systems the
classical description of the ``vacuum'' on either side of the horizon
fails on length scales smaller than a characteristic length $\xi$, a
quantum-mechanical quantity, that diverges at the horizon. The classical
equations remain exactly valid up up to the horizon, but only in context
of a special, unphysical order of limits.  In a real experiment done at
finite size the diverging length $\xi$ will eventually reach this size and
cause the classical description of the experiment to fail. In the bose
fluid the approach to the critical surface is signaled by the vanishing of
the speed of sound.  In a black hole the approach to the horizon is
signaled by the vanishing of the time dilation factor.  An apt analogy
between the two thus requires the time dilation to {\it increase} inside
the event horizon--at odds with the prediction of classical general
relativity.

\begin{figure}
\epsfbox{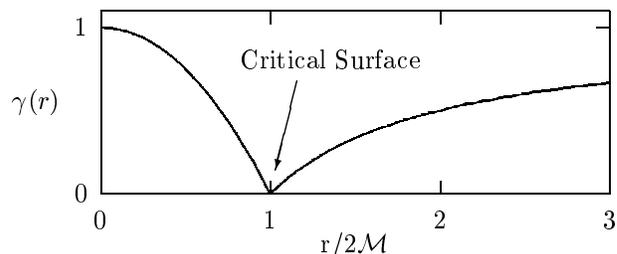}
\caption{Prototype time dilation factor $\gamma(r)$ in the vicinity
        of a black hole event horizon.}
\end{figure}

The notion that general relativity might be an emergent property in a
condensed-matter-like quantum theory of gravity has a long history
\cite{volovik}.  In 1968 Sakharov observed that space-time in Einstein
gravity was similar to to stressed matter \cite{sakharov}.  In Sakharov's
model there is no order parameter similar to that in superfluids, but one
with such order parameters has been proposed \cite{george}. In 1982 Unruh
observed a close analogy between sound propagation in background
hydrodynamic flow and field propagation in curved space-time \cite{unruh}. 
Following Unruh's lead, models for a black hole based on superfluid flow
of $^3$He \cite{jackobson} and atomic bose condensates \cite{garay} have
been put forward.  Mohazzab has recently proposed an analogy between black
hole event horizons and the normal-superfluid interface \cite{mohazzab} of
$^4$He. Ueda and Huang have noted the similarity between black hole
collapse and the instability of atomic bose condensates to attractive
forces \cite{ueda}.

      However, our proposal differs from this recent work in the key
respect that it ascribes black hole behavior at the event horizon to a
quantum ground state. This enables us to argue for the first time that
collective effects are the {\it correct} explanation for the puzzling
behavior of black holes--and by implication the apparent incompatibility
of quantum mechanics and general relativity.

\section{Bosonic Matter}

The simplest kind of matter is $^4$He and substances like it \cite{sound}. 
It is a collection of $N$ atoms obeying Eq. (1) with

\begin{equation}
{\cal H} = - \frac{\hbar^2}{2M} \sum_{j = 1}^N \nabla_j^2
+ {\cal V}({\bf r}_1 , ... , {\bf r}_N ) \; \; \; ,
\label{ham}
\end{equation}

\noindent
subject to the condition that $\Psi$ be symmetric under interchange of
any two of its arguments ${\bf r}_1 , ... , {\bf r}_N$. The ground
state is the energy eigenstate

\begin{equation}
\Psi({\bf r}_1 , ... , {\bf r}_N , t) = e^{- i E_0 t/\hbar} \;
\Phi({\bf r}_1 , ... , {\bf r}_N)
\end{equation}

\begin{equation}
{\cal H} \Phi = E_0 \Phi \; \; \; ,
\end{equation}

\noindent
with the lowest eigenvalue $E_0$.  The zero-temperature equation of state
of the matter is the functional dependence of $E_0$ on various parameters
in the hamiltonian, such as the confinement volume $V$ or atomic mass $M$.
This dependence is smooth and continuous except at quantum phase
transitions, where it is singular. 

The properties of $^4$He demonstrate that both zero-temperature phase
transitions of bosonic matter and the liquid phase exist \cite{helium}. 
$^4$He is a solid at pressures above 25 atmospheres and zero temperature. 
As the pressure is dropped below 25 bar it melts into a liquid with
unmeasurably small quantum vapor pressure--meaning that it puddles at the
bottom of a container larger than itself and will not evaporate at zero
temperature.  This liquid, like all the fluids we will consider here, is
much colder than the bose-einstein condensation temperature and is thus a
pure superfluid. 

The vapor phase of bosonic matter also exists in nature in the
newly-discovered ``bose-einstein condensates''--a name that is somewhat
misleading as these systems exhibit a finite sound speed \cite{sound}. 
They are also metastable excited states rather than ground states, and are
more aptly called supersaturated quantum vapors.  Their behavior is fully
consistent with Bogoliubov's original description of superfluid broken
symmetry in $^4$He, which was based on weak repulsive potentials and was
actually a description of the quantum gas \cite{bog}. 

The nature of the zero-temperature liquid-vapor transition in these
systems is, however, controversial.  In 1977 Miller, Nosanow and Parish
\cite{nosanow} performed a realistic variational study of lennard-jones
fluids and found that the critical point could not be reached by varying
pressure.  They concluded from this that bose fluids never have a
conventional critical point.  However there is no general principle
leading to that conclusion, and more recent studies based on different
model assumptions \cite{threebody} find behavior more consistent with that
of classical fluids. We will proceed on the assumption that the models
predicting a conventional critical point were solved correctly, and that
the result of Miller, Nosanow and Parrish was specific to the class of
model they were studying.

The quantum liquid-vapor transition, or something like it, may have been
seen experimentally in these condensates \cite{ketterle,wieman}.  The
relevant experiments exploit hyperfine scattering resonances in certain
isotopes to tune the s-wave phase shift through zero by means of a
magnetic field. When Cornish {\it et al.} \cite{wieman} did this with
$^{85}$Rb they found the ball of vapor to first contract - as expected if
its pressure were being reduced - and then explode partially, leaving a
remnant condensate with a ``halo'' of hot gas.  Interpretation of this
effect as a phase transition is complicated by the metastable nature of
the condensate and increased rate of recombination into the true ground
state that occurs at high densities.  However it occurs abruptly at the
place where such a transition is expected and is preceded by a dramatic
softening of the compressibility.

\begin{figure}
\epsfbox{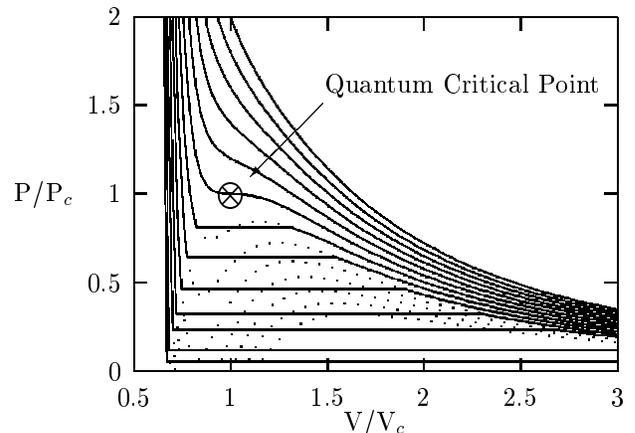}
\caption{Phenomenological equation of state defined by Eq. (\ref{eos})
         for various values of the parameter $c$ near the
         critical value.  The dotted lines indicate the Maxwell
         loops.}
\end{figure}

\section{Model Hamiltonian}

Let us now proceed to construct a model for the quantum liquid-vapor
transition. The simplest realization of this transition in a
classical fluid is the Van der Waals equation of state \cite{waals}

\begin{equation}
(V - b)(P + \frac{a}{V^2}) = N k_B T \; \; \; .
\end{equation}

\noindent
By analogy with this let us consider the phenomenological quantum
equation of state

\begin{equation}
(V^2 - b) (P + \frac{a}{V^4}) = c \; \; \; ,
\label{eos}
\end{equation}

\noindent
shown in Fig. 2.  This is generated in the mean-field approximation from
the field theory

\begin{equation}
{\cal L} = \psi^* ( i \hbar \frac{\partial}{\partial t} + \mu ) \psi -
\frac{\hbar^2}{2M} | {\bf \nabla} \psi |^2 -
{\cal U}(|\psi |^2) \; \; \; ,
\label{lagrange}
\end{equation}

\noindent
with $| \psi|^2$ interpreted as the density $\rho = N/V$ and

\begin{equation}
{\cal U} = \frac{c}{2 \sqrt{b} V} \ln
( \frac{V + \sqrt{b}}{V - \sqrt{b}})
- \frac{a}{3 V^4} \; \; \; .
\end{equation}

\noindent
After canonical quantization this becomes equal to Eq. (\ref{ham}) with a
short-range multiconfigurational potential ${\cal V}$.  At zero temperature
this system exhibits the phenomenon of bose condensation--i.e. acquires
a superfluid order parameter $\psi$ with low-energy dynamics
described by the extremal condition

\begin{equation}
i \hbar \frac{\partial \psi}{\partial t} = - \frac{\hbar^2}{2M}
\nabla^2 \psi + [{\cal U}' (|\psi |^2 ) - \mu] \psi
\; \; \; .
\end{equation}

\noindent
This is the Gross-Pitaevskii equation \cite{pitaevski}. The particle
density and current density, defined by

\begin{equation}
| \psi |^2 = \rho
\; \; \; \; \; \; \;
\rho {\bf v} = \frac{\hbar}{2Mi} ( \psi^* {\bf \nabla} \psi -
\psi {\bf \nabla} \psi^* ) \; \; \; ,
\end{equation}

\noindent
then satisfy hydrodynamic conservation of particle number and momentum
\cite{landau}

\begin{equation}
\frac{\partial \rho}{\partial t} + {\bf \nabla} \cdot ( \rho {\bf v}) = 0
\; \; \; \; \; \;
M \frac{\partial}{\partial t} ( \rho {\bf v} ) + {\bf \nabla} P
= 0 \; \; \; .
\label{hydro}
\end{equation}

\noindent
The quiescent state of the fluid is described by the uniform solution
$\psi_0$ satisfying

\begin{equation}
{\cal U}' (|\psi_0 |^2 ) - \mu \psi_0 = 0
\; \; \; \; \; \; \; \;
(P V + E_0 = \mu)
\; \; \; .
\end{equation}

\noindent
The particle density is fixed by suitably adjusting the chemical
potential $\mu$. Small perturbations to this solution

\begin{equation}
\psi = \psi_0 + \delta \psi_{\rm R} + i \; \delta \psi_{\rm I}
\end{equation}

\noindent
then satisfy

\begin{equation}
\hbar \frac{\partial (\delta \psi_R )}{\partial t}
= - \frac{\hbar^2}{2M} \nabla^2 (\delta \psi_{\rm I} )
\label{psir}
\end{equation}

\begin{equation}
- \hbar \frac{\partial (\delta \psi_I)}{dt} =
- \frac{\hbar^2}{2M} \nabla^2 (\delta \psi_{\rm R} )
+ \frac{2 B}{\rho} (\delta \psi_{\rm R})
\; \; \; ,
\label{psii}
\end{equation}

\noindent
to linear order and thus give the dispersion relation

\begin{equation}
\hbar \omega_q = \sqrt{ (\hbar v_s q)^2 + (\frac{\hbar^2q^2}{2M})^2}
\label{dispersion}
\end{equation}

\noindent
for compressional sound.  This identifies $\xi = \hbar / M v_s$ as the
length scale for the failure of hydrodynamics.  This same scale appears
is the Bogoliubov solution \cite{bog}.

As usual the region of negative compressibility is an inaccurate
description of liquid-gas phase separation and is replaced with a Maxwell
construction.  This is discussed more thoroughly in Appendix A.  For this
reason there is one and only one point in the diagram where the bulk
modulus $B = - V (\partial P / \partial V)$ is zero, namely the critical
point.

\begin{figure}
\epsfbox{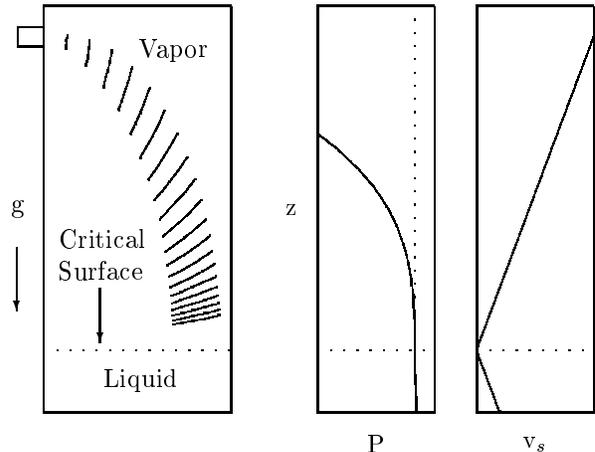}
\caption{Illustration of thought experiment in which pressure increases
        toward the bottom of a tank of quantum fluid.  Sound emitted from
        a transducer on the side of the tank is refracted downward toward
        the critical surface where the sound speed collapses to zero.  The
        wave fronts shown are for a solution of Eq. (\ref{wave}) with the
        pressure and sound speed profiles given by Eq. (\ref{profile}) and
        plotted on the right.  The quantum pathologies in this case are
        exactly the same as those at a Schwarzschild black hole.}
\end{figure}

\section{Critical Surface Event Horizon}

Let us now imagine a thought experiment, illustrated in Fig. 3, in which a
tall tank on the surface of the earth is filled with a quantum fluid
characterized by a critical equation of state. The pressure increases
toward the bottom of the tank due to gravity and at some critical depth
reaches, and then surpasses, the critical pressure. Sound waves are
refracted toward this surface just as light is refracted toward a
black hole horizon and for the same reason, namely that the propagation
speed measured by a clock at infinity vanishes there. For the specific
equation of state defined by Eq. (\ref{eos}) with $c = 8 a / 27 b$,
which reduces near the critical point to

\begin{equation}
\rho_c = \frac{1}{\sqrt{3b}}
\; \; \; \; \; \;
P_c = \frac{a}{27 b^2}
\; \; \; \; \; \; \; \; \; \;
v_c = \sqrt{\frac{P_c}{M \rho_c}}
\end{equation}

\begin{equation}
\frac{P}{P_c} - 1 \simeq 12 \; ( \frac{\rho}{\rho_c} - 1 )^3
\; \; \; ,
\end{equation}

\noindent
we have

\begin{equation}
\frac{P}{P_c} \simeq 1 - \frac{g z}{v_0^2}
\; \; \; \; \; \;
\frac{v_s}{v_0} \simeq 6 \biggl|
\frac{gz}{12 v_0^2} \biggr|^{1/3}
\; \; \; .
\end{equation}

\noindent
near the critical surface at $z = 0$. Small density fluctuations $\rho
\rightarrow \rho + \delta \rho$ then propagate according as

\begin{equation}
{\bf \nabla} \cdot [ v_s^2 \; {\bf \nabla} (\delta \rho ) ] =
\frac{\partial^2 (\delta \rho )}{\partial t^2}
\; \; \; .
\label{wave}
\end{equation}

\noindent
This is qualitatively the same as the scalar wave equation

\begin{equation}
{\bf \nabla} \cdot [ v_s \; {\bf \nabla} \phi ] = \frac{1}{v_s}
\frac{\partial^2 \phi}{\partial t^2}
\end{equation}

\noindent
one obtains from

\begin{equation}
\frac{\partial}{\partial x^\mu} ( \sqrt{-g} \; g^{\mu \nu}
\frac{\partial \phi}{\partial x^\nu} ) = 0
\end{equation}

\noindent
using the gravitational metric

\begin{equation}
ds^2 = g_{\mu \nu} dx^\mu dx^\nu = dx^2 + dy^2 + dz^2 - v_s^2 dt^2
\; \; \; .
\end{equation}

\noindent
The particular power law with which the sound speed vanishes in this
experiment is not important and can easily be modified.  For example
one can imagine weakening the downward force on the atoms according to the
rule

\begin{equation}
g = g_0 ( 1 - e^{- z^2/\ell^2} ) \; \; \; ,
\end{equation}

\noindent
so that

\begin{equation}
\frac{P}{P_c} \simeq 1 - \frac{g_0 z^3}{3 \ell^2 v_0^2}
\; \; \; \; \; \;
\frac{v_s}{v_0} \simeq 6 \biggl|
\frac{g_0 z^3}{36 \ell^2 v_0^2} \biggr|^{1/3} \; .
\label{profile}
\end{equation}

\noindent
The analogy with gravity is more obvious in this case because the metric
just outside the event horizon of a Schwarzschild black hole can be
written in this form with $v_s = c^3 z / 4 G M$.

\section{Quantum-Critical Dissipation}

In contrast to the case of classical gravity, however, the paradoxes of
sound propagation near the critical surface have a simple
quantum-mechanical resolution:  Sound ceases to make sense near the
horizon because the principles of hydrodynamics fail on length scales
smaller than the correlation length $\xi = \hbar / M v_s$ and time scales
longer than $\xi / v_s$. At the horizon both diverge to infinity. A sound
quantum with fixed frequency $\omega$ propagating toward the horizon
reaches the point at which $\omega \geq v_s/\xi$ in finite time and decays
there into the soft excitations of the critical point.  These are dense,
so most of the energy thermalizes.  This effect has never been observed
experimentally, but its classical analogue, critical opalescence, is well
known and has been studied extensively by light scattering \cite{opal}.

Let us now consider this effect in detail.  At the critical point the
Lagrangian is effectively

\begin{equation}
{\cal L}_{\rm eff} = \psi^* ( i \hbar \frac{\partial}{\partial t}
- \mu) \psi - \frac{\hbar^2}{2M} | {\bf \nabla} |^2
- \frac{3 P_c}{\rho_c^2} (| \psi | - \psi_0 )^4
\end{equation}

\noindent
i.e. a nonrelativistic bose gas with a high-order nonlinearity.  The
corresponding quantum Hamiltonian is

\begin{displaymath}
{\cal H}_{\rm eff} = \sum_{\bf q} \frac{\hbar^2 q^2}{2M}
a_q^\dagger a_q + \frac{3 P_c \psi_0^4}{V \rho_c^2} \sum_{{\bf q}_1
{\bf q}_2 {\bf q}_3 {\bf q}_4 }
\end{displaymath}

\begin{displaymath}
\times \delta ({\bf q}_1 + {\bf q}_2 + {\bf q}_3 + {\bf q}_4 )
(a_{{\bf q}_1} + a_{-{\bf q}_1}^\dagger )
(a_{{\bf q}_2} + a_{-{\bf q}_2}^\dagger )
\end{displaymath}

\begin{equation}
\times (a_{{\bf q}_3} + a_{-{\bf q}_3}^\dagger )
(a_{{\bf q}_4} + a_{-{\bf q}_4}^\dagger ) \; \; \; .
\end{equation}

\begin{figure}
\epsfbox{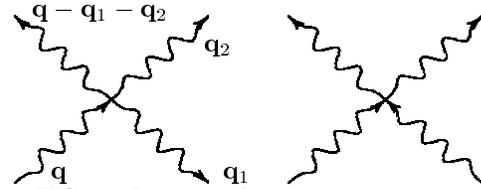}
\caption{Lowest-order scattering processes in the critical region.  The
        process on the left causes mass renormalization and decay at
        zero temperature. The one on the right causes critical
        opalescence.}
\end{figure}

\noindent
The important decay and scattering processes are shown in Fig. 4.  The
first correction to the the particle self-energy renormalizes $M$ and
gives an imaginary part

\begin{displaymath}
{\rm Im} \Sigma_{\bf q} (\omega) = ( \frac{3 P_c}{\rho_c^2 V})^2
\sum_{{\bf q}_1 {\bf q}_2 } {\rm Im} \biggl[
\hbar \omega
\end{displaymath}

\begin{displaymath}
- \frac{\hbar^2}{2M} (| {\bf q}_1 |^2 + | {\bf q}_2 |^2
+ | {\bf q}_1 + {\bf q}_2 + {\bf q} |^2 ) + i \eta \biggr]^{-1}
\end{displaymath}

\begin{equation}
= - \frac{3}{16 \pi^2} ( \frac{M}{\hbar^2})^3
( \frac{P_c}{\rho_c^2} )^2 (\hbar \omega - \frac{\hbar^2 q^2}{6M} )^2
\Theta ( \hbar \omega- \frac{\hbar^2 q^2}{6M} ) \; .
\end{equation}

\noindent
The decay rate for a boson of energy $\hbar \omega = \hbar^2 q^2 / 2 M$ is
thus

\begin{equation}
\frac{\hbar}{\tau} =\frac{1}{3\pi^2} (\frac{M}{\hbar^2} )^3
(\frac{P_c}{\rho_c^2})^2 (\hbar \omega)^2 \; \; \; .
\label{decay}
\end{equation}

\noindent
This implies that the free boson becomes more and more sharply defined
as the energy is lowered, so that in the low-energy limit one retrieves
the ideal noninteracting bose gas \cite{sachdev}.

Let us now consider the several experimental signatures of this effect:

\begin{figure}
\epsfbox{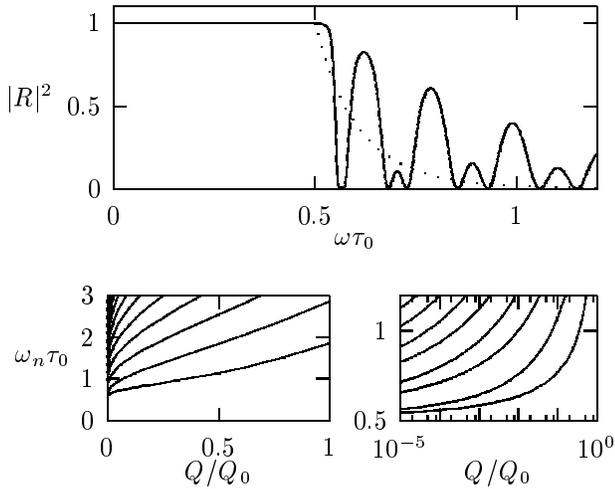}
\caption{Top: Reflectivity as a function of $\omega \tau_0$ predicted
        by Eq. (31) for the case of $Q/Q_0 = 10^{-5}$, with $Q_0 =
        (2M / \hbar \tau_0)^{1/2}$.  The interface thickness is assumed
        to be of order $1/Q_0$ to make the resonances visible. The
        dotted line is the $Q \rightarrow 0$ limit described by Eq.
        (\ref{normal}). Bottom:  Dispersion relation of interface bound
        states plotted both linearly (left) and semi-logarithmically
        (right).}
\end{figure}

\subsection{Reflectivity}

At very low frequencies a phonon impinging on the surface at zero
temperature is coherently reflected or transmitted depending on its
energy.  Solving the equation

\begin{equation}
\hbar^2 \frac{\partial^2 \phi}{\partial t^2} = 
(\frac{\hbar}{\tau_0} )^2
{\bf \nabla} \cdot ( z^2 {\bf \nabla} \phi ) - ( \frac{\hbar^2}
{2M} )^2 {\bf \nabla}^4 \phi \; \; \; ,
\label{reflect}
\end{equation}

\noindent
where $1 / \tau_0 = \partial v_s / \partial z$ ({\it cf.} Eqs.
(\ref{psir}) and (\ref{psii}) with $\delta \psi_{\rm R} = \partial
\phi / \partial z$), we obtain the reflection coefficient shown in Fig. 5. 
The momentum component $Q$ in the plane, which is conserved, acts like a
mass and allows the phonon to become trapped at certain energies $\hbar
\omega_n$. These produce transmission resonances that become narrower and
narrower with increasing thickness of the interface.  The positions of
these resonances depend on $M$ and may thus be used spectroscopically to
determine this parameter.  For $Q >> Q_0$ they occur at the harmonic
oscillator values

\begin{equation}
\hbar \omega_n \simeq \frac{\hbar^2 Q^2}{2M} + ( n + \frac{1}{2} )
\sqrt{2} \frac{\hbar}{\tau_0} \; \; \; .
\end{equation}

\noindent
Normal incidence ($Q \rightarrow 0$) is a singular point where the
discrete energies collapse to a continuum characterized by the
reflectivity

\begin{equation}
| R |^2 = \left[ \begin{array}{ll}
1 & \omega \tau_0 < 1/2 \\
\cosh^{-2}( \pi \sqrt{(\omega \tau_0)^2 - 1/4}) & \omega \tau_0 > 1/2
\end{array} \right]
\; \; \; .
\label{normal}
\end{equation}

\noindent
This result is discussed further in Appendix B.

\begin{figure}
\epsfbox{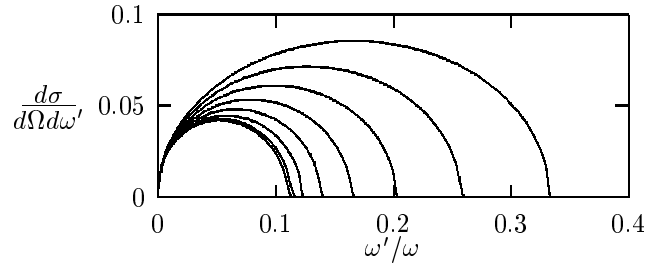}
\caption{Differential cross-section given by Eq. (\ref{cross}) for
        inelastic scattering of sound from a critical surface as a
        function of scattered frequency $\omega'$ for values of $\theta$
        ranging from 0 to $\pi/2$.  The maximum value of $\omega'/\omega$
        for sound reflected normally $(\theta = 0)$ is $1/9$.}
\end{figure}

\subsection{Inelastic Scattering}

The horizon is opaque to high-frequency sound waves impinging upon it and
inelastically scatters about 1/8 of them back out with a strong red shift.
When $\tau_0/\tau > 1$, where $\tau$ is given by Eq. (\ref{decay}),
incoming phonon decays with 100\% probability, and one (but not two) of
the three bosons thus generated can escape back out the surface. This
gives a differential cross section per unit area $A$ to scatter sound of
frequency $\omega$ back in solid angle $d\Omega$ at frequency $\omega' <
\omega$ of

\begin{equation}
\frac{d \sigma}{d \Omega d\omega'} = \frac{27 A}{16 \pi^2 \omega}
\sqrt{3 x [ 1 - 3 x - 2 \sqrt{x} \cos(\theta) ] } \; \; \; ,
\label{cross}
\end{equation}

\noindent
where $x = \omega ' / \omega$.  This is plotted in Fig. 6.  Thus the
horizon is a blackbody that fluoresces red.  It is an extremely efficient
thermalizer of energy, however.

\begin{figure}
\epsfbox{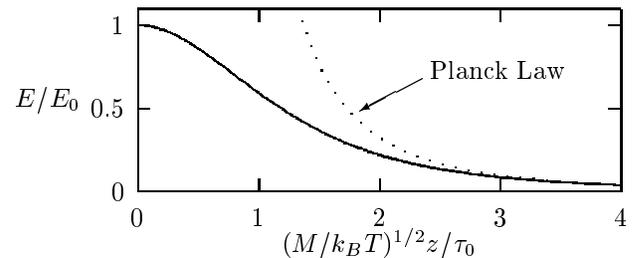}
\caption{Thermal energy per unit volume defined by Eqs. (\ref{heat}) and
        (\ref{edense}) for a critical surface at temperature $T$.  The
        divergent planck law is shown for comparison.}
\end{figure}

\subsection{Heat Capacity}

The horizon becomes much more dissipative at low frequencies if the
critical region is hot. At temperatures below the bose condensation
temperature ($k_B T_B \simeq \hbar^2 \rho^{2/3}/2M$) the absorption rate
for a phonon of energy $E << k_B T$ is roughly Eq. (\ref{decay}) with $k_B
T$ substituted for $E$ ({\it cf.} Fig. 4). This effect is equivalent to
classical critical opalescence. Its heat capacity is large but finite.
The energy density a distance $z$ away from the interface is

\begin{equation}
\frac{E}{V} = \frac{1}{2\pi^2} \int_0^\infty \frac{\hbar \omega_q \; q^2
dq} {\exp(\beta \hbar \omega_q) - 1} \; \; \; ,
\label{heat}
\end{equation}

\noindent
where $\omega_q$ is given by Eq. (\ref{dispersion}) with $v_s = z /
\tau_0$.  This is plotted in Fig. 7.  It may be seen to limit properly as
$z \rightarrow 0$ to

\begin{equation}
\frac{E_0}{V} = 0.128 \; (\frac{M}{\hbar^2})^{3/2} (k_B T )^{5/2}
\; \; \; .
\label{edense}
\end{equation}

\noindent
Thus the criticality cuts off the divergence in the planck law $E/V =
(\pi^2/30) (k_B T)^4 (\tau_0/\hbar z )^3$.

\section{Dimensional Analysis}

It is unfortunately not the case that knowledge of the mechanical
analogues of $c$ and $G$ is sufficient to determine the correlation
lenth $\xi$. The fluid possess a dimensionless parameter $\eta = \xi^3
\rho$ that cannot be determined by any low-frequency measurement.  There
is indeed a unique combination of the sound speed $v_s$, mass density $M
\rho$, and $\hbar$ that has units of length, but this cannot be associated
with $\xi$ without assuming that $\eta = 1$, which need not be the case. 

The analogue of Newton's constant $G$ in the fluid is the inverse mass
density $1/M\rho$.  There is no inverse-square attraction between two
masses in the superfluid because its monopolar nature makes this
interaction short-ranged.  However, we can sensibly compare the radiation
produced by rotating quardupoles. A pair of masses ${\cal M}$ orbiting
around each other at separation $\ell$ at frequency $\omega_0$ radiate
gravitational power

\begin{equation}
{\cal P} = \frac{2}{15} \frac{G}{c^5} {\cal M}^2 \ell^4 \omega_0^6
\; \; \; .
\end{equation}

\noindent
Two analogous masses in the fluid radiate sound power

\begin{equation}
{\cal P} = \frac{1}{15\pi} \frac{1}{v_s^5 M \rho} {\cal M}^2
\ell^4 \omega_0^8 \; \; \; .
\label{power}
\end{equation}

\noindent
These masses are polarons formed around an extremely light impurity, such
as an electron in $^4$He \cite{polaron}. This result is discussed in more
detail in Appendix C.

By analogy, then, the correlation length of gravity need not be the planck
length $\xi_p = (\hbar G/c^3)^{1/2}$.

\section{Interior Metric}

Let us now consider the implications of this analogy for real gravity.
While it is generally impossible to infer the properties of the second
phase from measurements made outside the black hole, the simplest guess is
that it is literally like the liquid-vapor transition, meaning that the
Einstein field equations, like the laws of quantum hydrodynamics, are
valid in both phases. Thus we have the following constraints:

\begin{enumerate}

\item The equations of classical general relativity outside the black hole 
      are obeyed everywhere except the critical surface.

\item At the critical surface the vacuum of space-time reorganizes itself
      so as to keep global time defined.

\item The local properties of the vacuum just inside the critical
      surface are indistinguishable from those just outside. 

\item There are no scales other than the mass.

\item The topology is consistent with the collapse of ordinary
      matter.

\end{enumerate}

\noindent
The physical indistinguishability of the quantum fluids on either side of
a liquid-vapor critical surface is a strong constraint on any
gravitational analogue because it requires the relativity principle to
operate on both sides.  This, in turn, requires that a metric be defined
and obey equations something like the Einstein field equations on both
sides.  The considerations at the critical surface then extrapolate to the
entire bulk interior, since an inability to do so would imply a second
phase boundary. If the metric exists in the interior of the black hole
then one can measure its curvature and compute from this the Einstein
tensor. This must be {\it real} stress-energy, because if it is not then
the space-time has a local property distinguishing it from the space-time
outside. 

Let us now write these ideas formally. The most general spherically
symmetric metric is \cite{eddington}

\begin{equation}
ds^2 = e^\lambda dr^2 + r^2 [ d\theta^2 + \sin^2 (\theta) d\phi^2 ]
- e^\nu dt^2 \; \; \; .
\end{equation}

\noindent
The corresponding stress-energy tensor is

\begin{equation}
R_{11} - \frac{1}{2} g_{11} R = - \nu' / r - ( 1 - e^\lambda)/r^2
\label{r11}
\end{equation}

\begin{displaymath}
R_{22} - \frac{1}{2} g_{22} R
= \frac{1}{\sin^2 (\theta )} \biggl[
R_{33} - \frac{1}{2} g_{33} R \biggr]
\end{displaymath}
 
\begin{equation}
= - r^2 e^{-\lambda} 
( \frac{\nu''}{2} - \frac{\lambda' \nu'}{4}
+ \frac{\nu'^2}{4} + \frac{\nu' - \lambda '}{2 r} )
\label{r22}
\end{equation}

\begin{equation}
R_{00} - \frac{1}{2} g_{00} R = e^{\nu - \lambda} [
- \lambda ' / r + ( 1 - e^\lambda)/r^2 ]
\; \; \; .
\label{r00}
\end{equation}

\noindent
Outside the black hole the Einstein equations require this to be
zero, which gives the Schwarzschild solution $\gamma(r) = e^\nu =
e^{-\lambda} = 1 - 2{\cal M}/r$.  The choice of integration constant
$2{\cal M}$ determines the location of the event horizon. If we then
require the horizon be a critical surface we must also have $e^\nu = e^{-
\lambda}$ immediately inside the horizon as well, but converging to zero
with the opposite slope.  This, in turn, requires the presence of matter
with negative pressure inside the black hole. ($\nu'$ and $1 - e^\lambda$
are both negative at the horizon but must limit to zero at the origin.)
The matter must also naturally resist falling into the minimum of the
gravitational potential, which necessarily lies at the horizon.  These
properties are so difficult to achieve with any kind of conventional
matter the only reasonable choice is a nonzero cosmological constant.
Thus inside the black hole we must have

\begin{equation}
R_{\mu \nu} - \frac{1}{2} g_{\mu \nu} R = \frac{3}{4{\cal M}^2} \;
g_{\mu \nu} \; \; \; ,
\end{equation}

\noindent
where the constant ${\cal M}$ is picked to match the boundary condition
at $r = 2 {\cal M}$. This solution has the additional useful feature that
the energy inside the black hole sums correctly to ${\cal M}$.  The result
is the metric

\begin{equation}
\gamma (r) = \left[ \begin{array}{ll}
1 - (r/2{\cal M} )^2 & r < 2{\cal M} \\
1 - 2{\cal M}/r & r > 2 {\cal M} \end{array} \right]
\label{gamma}
\end{equation}

\noindent
shown in Fig. 1.  It corresponds physically to a vacuum vessel containing
a region of space-time with a positive cosmological constant. Note the
similarity to Fig. 3.
 
The singularity at the event horizon corresponds to a negative surface
tension or stress required to contain the negative pressure inside the
black hole.  It may be seen from Eqs. (\ref{r11}) - ( \ref{r00}) to show
up only in the $22$ and $33$ components of the Einstein tensor.  A balloon
with surface tension ${\cal T}$ filled gas at pressure $P$ will acquire a
radius $r$ satisfying ${\cal T} = r P / 2$.  Similarly the black hole with
local pressure $P = - 3 c^8 / 32 \pi G^4 {\cal M}^2$ in proper coordinates
inside must have surface tension ${\cal T} = - 3 c^2 / 32 \pi G^2 {\cal
M}$ at the horizon in proper coordinates.  This tension is generated by
the space-time itself as it undergoes the transition between its two
phases and thus need not be constrained by the properties of any familiar
kinds of matter. However, it is actually quite small. To see this let us
imagine emulating the stress by generating thermal photons at infinity and
allowing them to fall down on the black hole.  The light pressure in
proper coordinates at the horizon is formally divergent because of the
gravitational potential.  However, this is false pathology because proper
coordinates do not make physical sense at the horizon. Per Fig. 7, the
light pressure measured in proper coordinates far away from the black hole
is actually finite.  The cosmological constant pressure in these same
coordinates is zero at the horizon. It is thus always negligible compared
to any background thermal radiation pressure. 

If the event horizon is indeed a critical surface then its heat capacity
measured by distant observers is finite. With $\omega_q$ defined as in Eq.
(\ref{dispersion}) with $v_s = c \gamma^{1/2} (r)$ we have for the the
total energy per bosonic degree of freedom inside the event horizon

\begin{displaymath}
E = \int_0^{2G{\cal M}/c^2} \biggl[ \frac{\gamma^{-1/2}(r)}{2\pi^2}
\int_0^\infty q^2 \; \frac{\hbar \omega_q}{\exp (\beta \hbar \omega_q )
- 1} \; dq \biggr] r^2 dr
\end{displaymath}

\begin{equation}
\simeq 1.1 \times \biggl\{ \biggl[ \frac{4\pi}{3} ( \frac{2 G {\cal M}}
{c^2} )^3 \biggr] \; \biggl[ \frac{\pi^2}{30} \frac{(k_B T)^4}
{(\hbar c)^3} \biggr] \biggr\} \frac{Mc^2}{k_B T}
\; \; \; .
\label{capacity}
\end{equation}

\noindent
Thus the heat content of the black hole is $\sim Mc^2 / k_B T$ times
the volume of empty space of the same radius.  This may also be written
$E = 7.6 \times 10^{-4} \; Mc^2 (T / T_H )^3$, where $T_H = \hbar c^3 / 8
\pi k_B G {\cal M}$ is the Hawking temperature.

\section{Discussion}

The resolution of the black-hole paradox we have proposed here conflicts
fundamentally with the relativity principle, in that it requires quantum
gravity to have a mass scale $M$ that can be measured.  If such a scale
exists at a black hole horizon then it must exist in asymptotically flat
space-time as well and correspond to an absolute velocity scale at which a
particle gains mass and loses integrity.  This is not so different from
the effects of a new elementary particle at this scale, except that decays
normally forbidden by relativistic kinematics, i.e.  one photon going to
three, become possible. No such scale has ever been observed.  However the
idea that Einstein gravity is emergent in the sense we describe is
inherently falsifiable.  The relativity principle itself must break down
at sufficiently high energy scales, and this breakdown must show up
experimentally as spontaneous decay of bosons, such as photons, that
otherwise should have integrity.  This might have observable effects on
the highest-energy cosmic rays. 

Our theory also predicts that black holes have specific spectroscopic
signatures that can be observed from outside the horizon. By analogy with
Fig. 5 we expect the horizon to be highly reflective to light of frequency
less than c times the black hole radius and to transmit light slightly
above this frequency in resonances that depend on the angle of incidence. 
In terms of the planck mass $M_p = (\hbar c / G)^{1/2} = 2.18 \times
10^{-5}$ gm and the mass of the sun $M_\odot = 2 \times 10^{33}$ gm we
have

\begin{equation}
\tau_0 = \frac{ 2 G {\cal M}}{c^3} =
(\frac{{\cal M}}{M_\odot}) \times  1.00 \times 10^{-5} \;
{\rm sec}
\end{equation}

\begin{equation}
Q_0 = \sqrt{\frac{2M}{\hbar \tau_0}} =
\sqrt{(\frac{M}{M_p})(\frac{M_\odot}{{\cal M}})}
\times  6.47 \times 10^{13} \; {\rm cm}^{-1} \; .
\end{equation}

\noindent
Thus if $M$ is comparable to the planck mass then the transverse momentum
$Q$ of an incoming photon will always be small compared with $Q_0$ unless
its energy far from the black hole exceeds $\hbar c Q_0 = 1.28 \times
10^9$ eV.  This implies that the reflectivity of Fig. 5 is a fairly apt
description of what one would see for a cold solar-mass black hole.  Both
the reflection threshold and the transmission resonances would be in the
radio near $10^5$ sec$^{-1}$. The reflectance edge is similar to a
classical effect caused by the convergence of the radial coordinate
\cite{price}, but the resonances have no classical analogue. By analogy
with the inelastic scattering of high-frequency sound from a critical
surface, we expect that high-frequence electromagnetic radiation will be
inelastically backscattered from the event horizon with a characteristic
spectrum terminating at a red shift of 90\% for normal incidence. Exactly
how high the energy must be for this process to be efficient depends on
the matrix element for decay.  If we assume the latter to be set by planck
units also, then we have $\tau_0 / \tau \sim (E / \hbar c Q_0 )^2$ .  This
implies that the process is efficient only for hard gamma rays, and that
most photons with energies less than $10^9$ eV pass through the horizon
without decaying.  Once on the other side they refract away from the
center of the black hole, as in a defocussing lens.

We also expect the black hole horizon to be a thermalizer of radiation and
to be itself a thermal body with a finite positive heat capacity measured
by an observer at infinity.  This heat capacity is comfortably small for
astrophysical objects.  Rewriting Eq. (\ref{capacity}) as

\begin{equation}
\frac{E}{{\cal M} c^2} = 3.24 \times 10^{-20} \;
(\frac{M}{M_p}) (\frac{{\cal M}}{M_\odot})^2 (\frac{T}{1 ^\circ K})^3
\; \; \; ,
\end{equation}

\noindent
we that that if $M$ is the planck mass then the heat content of a
solar-mass black hole becomes comparable to its mass when $T \sim 10^6$
$^\circ$K.  This implies that the temperature of a solar-mass black hole
might well be sufficiently high ($> 10^3$ $^\circ$K) to make it visible
against the cosmic microwave background. 

The specific metric we propose identifies the second phase as de Sitter
space and its distinguishing characteristic as a nonzero cosmological
constant.  However, it is arguable that the key distinguishing
characteristic is not the cosmological constant, per se, but topology. The
cosmological constant we find depends on the black hole mass and becomes
unmeasurably small when the latter is large. The horizon is effectively
planar in this limit, and the two phases locally indistinguishable.
However, one can see from Figs. (2) and (3) that exactly the same thing
occurs at the liquid-vapor critical point.  The liquid and vapor sides of
the transition are distinguished only in how the critical equation of
state eventually deviates from symmetric inflection, which is a global
property.  However, one can imagine resolving this problem by eliminating
the earth's gravitational field in Fig. 3 and substituting the field due
to self-gravitation of the fluid.  Then ``down'' is determined by the
center of gravity of the fluid, the critical surface is a sphere, and the
two phases are distinguished as a practical matter by which is inside the
sphere and which outside. 

While we cannot rule out on any technical grounds the possibility that the
transition is first-order, we find it highly unlikely because it would
require the the metric to be discontinuous. Referring to Fig. 2, we see
that perturbing the critical equation of state downward causes the
susceptibility of the soft excitation - in this case sound - to become
negative, so that density perturbations of the uniform state grow.  The
uniform state is thus absolutely unstable to a nonuniform one
characterized by position-dependent density.  The relevant soft excitation
in Einstein gravity is a gravity wave

\begin{equation}
\delta g_{xx} = - \delta g_{yy} \sim e^{i ( k z - \omega t)}
\; \; \; .
\end{equation}

\noindent
If this excitation were to become unstable in the same way it would
generate a nonuniform metric with sharp jumps analogous to the density
jumps at liquid-vapor interfaces discussed in Appendix A. This would be a
much more violent breakdown of classical general relativity, and in
particular could not be interpreted as interface stress-energy. 

Regardless of whether it the event horizon corresponds to a first- or
second-order phase transition, identifying the space-time of a black hole
as a quantum ground state resolves the information paradox
\cite{chapline}. The horizon does not destroy quantum information but
rather makes entropy the same way black paint does, i.e. by scattering the
energy into a thermodynamically large number of degrees of freedom.  This
is conceptually similar to the quantum holography ideas \cite{susskind},
except that the relevant degrees of freedom are collective in nature
rather than fundamental. Also, we find that the that vacuum beyond the
horizon is locally identical to the one we know and can be probed
experimentally from the outside.  Insofar as string theory predicts
something else, the two theories can be distinguished from each other by
experiment. 

Our picture for black hole is also fundamentally different from the
classical one, in that we find quantum effects determine both the nature
of the event horizon and the interior spacetime, even in the case of
macroscopic black holes.  The possibilities for falsifying our predictions
and thereby demonstrating the quantum nature of black holes is, of course,
most exciting. 

\section*{Acknowledgements}

This work was supported primarily by the National Science Foundation
under Grant DMR-9813899.  Additional support was provided by NASA
under Collaborative Agreement No. NCC 2-794.

\appendix

\section{Maxwell Construction}

The Maxwell construction of the zero-temperature equation of state shown
in Fig. 2 comes from the condition that the superfluid order parameter be
extremal across the liquid-vapor interface.  As is the case at finite
temperature the end points of the Maxwell loops have identical pressures
$P$ and chemical potentials:

\begin{equation}
{\cal U} (\psi_1^2) - \mu \psi_1^2
= {\cal U} (\psi_2^2) - \mu \psi_2^2 = -P
\label{max1}
\end{equation}

\begin{equation}
{\cal U}' (\psi_1^2) = {\cal U}' (\psi_2^2) = \mu
\label{max2} \; \; \; .
\end{equation} 

\noindent
Let us now solve

\begin{equation}
- \frac{\hbar^2}{2M} \frac{d^2 \psi}{d z^2}
+ [ {\cal U}' (\psi^2) - \mu ] \psi = 0 
\end{equation}

\noindent
across the interface. Multiplying both sides of the equation by $d\psi /
dz$ we obtain

\begin{equation}
\frac{d}{dz} \biggl\{ - \frac{\hbar^2}{2M} ( \frac{d\psi}{dz} )^2
+ {\cal U} (\psi^2) - \mu \psi^2 \biggr\} = 0
\end{equation}

\begin{equation}
- \frac{\hbar^2}{2M} ( \frac{d\psi}{dz} )^2
+ {\cal U} (\psi^2) - \mu \psi^2 + P = 0
\; \; \; .
\label{c}
\end{equation}

\noindent
Integrating this by quadratures we then generate the density profile shown
in Fig. 8. Eqs. (\ref{max1}) and (\ref{max2}) guarantee that the profile
approaches $\psi_1$ and $\psi_2$ asymptotically. Deep in phase 1 we have,
with $\psi = \psi_1 + \delta \psi$,

\begin{equation}
{\cal U} ( \psi ) - \mu \psi^2 \simeq \frac{B_1}{2 \rho_1} \delta\psi^2
\; \; \; ,
\end{equation}

\noindent
and thus

\begin{equation}
\pm \xi_1 \int \frac{d (\delta \psi)}{\delta \psi} \simeq \int dz
\; \; \; \; \; \; \;
\psi - \psi_1 \sim e^{\pm z / \xi_1} \; \; \; .
\end{equation}

\noindent
The sign is picked to make $\delta \psi$ vanish away from the interface.  
Doing this for side 2 as well, we see that the thickness of the interface
region is essentially $\xi_1 + \xi_2$ and diverges at the critical point.

The surface tension is obtained by computing the total energy and
subtracting off the energy one would obtain in the limit of $\hbar
\rightarrow 0$ or $M \rightarrow \infty$, when the interface becomes sharp
and the energy comes entirely from ${\cal U}(\psi_1^2)$ and ${\cal
U}(\psi_2^2)$.  Let $\psi_0$ denote this solution.  Then the chemical
potential $\mu$ must be the same for $\psi_0$ as it is for $\psi$ since
the values of the two much match far away from the interface.  The
pressure $P$ is also be the same because neither $\psi_0$ nor
$\psi$ has a gradient away from the interface.  We also have

\begin{equation}
\mu \int (\psi^2 - \psi_0^2 ) dz = 0
\end{equation}

\noindent
because the number of particles in the two states is the same.  Now using
Eq. (\ref{c}) to eliminate ${\cal U}$ from the expression for the energy
difference, we obtain

\begin{equation}
{\cal T} =
\frac{\hbar^2}{M} \int (\frac{\partial \psi}{\partial z})^2 dz
\; \; \; .
\end{equation}

\noindent
Thus the surface tension is twice the kinetic energy per unit area.

\begin{figure}
\epsfbox{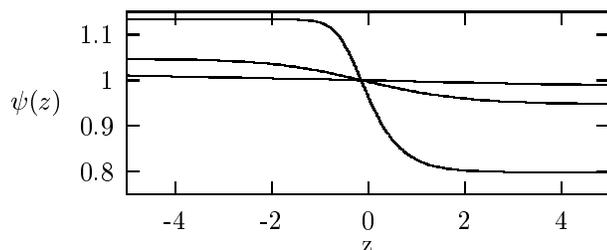}
\caption{Order parameter as a function of position across the
        liquid-vapor interface for the equation of state of Eq.
        (\ref{eos}) with the values 27bc/8a = 0.9, 0.99, and 0.999.  As
        the critical value 1 is approached the density jump across the
        interface collapses to zero and interface width diverges.}
\end{figure}

\section{Reflection Coefficient}

With momentum $Q$ in the plane of the interface, Eq. (\ref{reflect})
becomes

\begin{equation}
\omega^2 \phi = ( - \frac{\partial}{\partial z} z^2
\frac{\partial}{\partial z}
+ z^2 Q^2 ) \phi + ( - \frac{\partial^2}{\partial z^2} + Q^2 )^2 \phi
\; \; \; .
\end{equation}

\noindent
We eliminate the solutions of this equation that diverge as $\exp(z^2/2)$ 
by requiring $\phi(z)$ to have a fourier transform.  This satisfies

\begin{equation}
\omega^2 \hat{\phi} = \biggl[ - \frac{\partial}{\partial q} (q^2 + Q^2)
\frac{\partial}{\partial q} + (q^2 + Q^2 )^2 \biggr]  \hat{\phi}
\; \; \; .
\label{bound}
\end{equation}

\noindent
For $Q \neq 0$ this equation is regular at the origin and is easily solved
numerically.  The first 10 energy eigenfunctions for the case of $Q =
10^{-5}$ are shown in Fig. 9. As $Q$ becomes smaller and smaller the nodes
of the wavefunction are pulled into the origin. In the $Q \rightarrow 0$
limit the equation becomes singular at $q = 0$ and is no longer required
to be analytic there.  Thus we consider a wavefunction that is zero for $q
< 0$ and $\hat{\phi} = \sum_{n=0}^\infty a_n q^{n + \nu}$ for $q > 0$.
This satisfies the equation when

\begin{equation}
\nu = - \frac{1}{2} \pm \sqrt{1/4 - \omega^2}
\; \; \; \;
\frac{a_n}{a_{n-4}} = \frac{1}{n(n + 2 \nu + 1 )}
\; .
\end{equation}

\noindent
For $\omega < 1/2$ there is no normalizable solution.  For $\omega > 1/2$
on the other hand, there is always one normalizable solution formed by
subtracting the expressions for the two allowed values of $\nu$, namely

\begin{equation}
\hat{\phi} = (\frac{2}{q^2})^{1/4} \biggl[
I_{(2 \nu_1 + 1) / 4} (q^2/2) - I_{(2 \nu_2 + 1)/4} (q^2 / 2)
\biggr] \; .
\end{equation}

\noindent
Only the behavior of this function near $q \rightarrow 0$ is needed for
computing the reflectivity.  For large positive $z$ we have

\begin{displaymath}
\phi (z) \simeq \int_0^\infty
[ (\frac{q^2}{2})^{\nu_1 / 2} - (\frac{q^2}{2})^{\nu_2 / 2} ] \;
e^{i q z} \; dq
\end{displaymath}

\begin{equation}
= 2^{- \nu_1/2} \Gamma(\nu_1 + 1) |z|^{1 - \nu_1} e^{i \pi \nu_1 / 2} -
(1 \rightarrow 2) \; \; \; .
\end{equation}

\noindent
For negative z we have $\phi(z) = - \phi^* (-z)$.  The real and imaginary
parts of $\phi(z)$ are separately solutions, and we may combine them to
make a wave with no incoming component on the right side of the barrier. 
>From this we obtain the reflection and transmission coefficients

\begin{equation}
T = i 2^{i \alpha} \frac{\Gamma(1/2+ i \alpha)}
{\Gamma(1/2 - i \alpha)} \; \tanh
(\pi \alpha )
\end{equation}

\begin{equation}
R = 2^{i \alpha} \frac{\Gamma(1/2 + i \alpha)}
{\Gamma(1/2 - i \alpha)} \;
{\rm sech}( \pi \alpha)
\; \; \; ,
\end{equation}

\noindent
where $\alpha = \sqrt{\omega^2 - 1/4} \;$.

\begin{figure}
\epsfbox{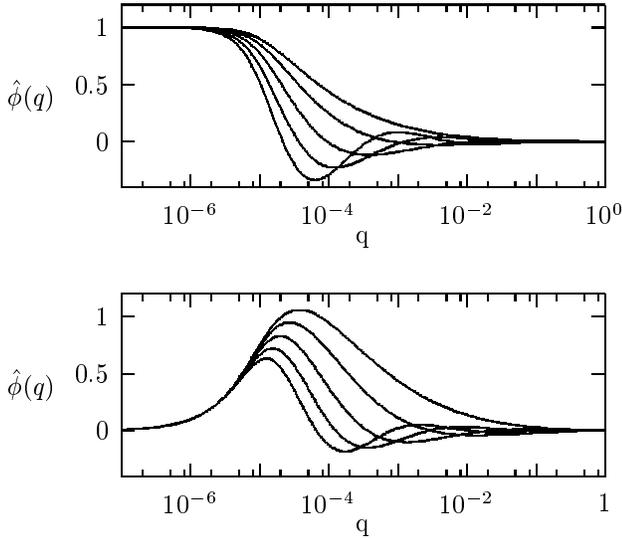}
\caption{Even (top) and odd (bottom) energy eigenstates of Eq.
        (\ref{bound}) for the case of $Q = 10^{-5}$.  The corresponding
        eigenvalue are shown in Fig. 5.}
\end{figure}

The resonant reflection spectrum one would actually measure is quite
sensitive to details of the experiment and cannot be computed without
further assumptions.  For the purposes of constructing Fig. 5 we assumed a
simple barrier with tunneling matrix elements increasing slowly with the
number of nodes in the wavefunction and a constant density of states at
infinity. Denoting this density of states, times the square of the scale
of the tunneling matrix elements, by $t$, we have

\begin{equation}
| R |^2 = 1 - \biggl| Tr \biggl[ t G^0 ( 1 + i t G^0 )^{-1} \biggr]
\biggr|^2 \; \; \; ,
\end{equation}

\noindent
where

\begin{equation}
G_{11}^0 = G_{22}^0 = \sum_n^N \frac{\sqrt{j+1}}
{(\omega - \omega_n + i \eta)}
\end{equation}

\begin{equation}
G_{12}^0 = G_{21}^0 = \sum_n^N (-1)^n \frac{\sqrt{j+1}}
{(\omega - \omega_n + i \eta)} \; \; \; .
\end{equation}

\noindent
Fig. 5 was generated using $t = 0.01$ and $N = 20$.

\section{Polaron Radiation}

The linearized equations of motion for $\psi = \psi_0 + \delta
\psi_{\rm R} + i \delta \psi_{\rm I}$ in the presence of a weak potential
$V({\bf r}, t)$ are

\begin{equation}
\hbar \frac{\partial (\delta \psi_R )}{\partial t}
= - \frac{\hbar^2}{2M} \nabla^2 (\delta \psi_{\rm I} )
\end{equation}

\begin{equation}
- \hbar \frac{\partial (\delta \psi_I)}{dt}
= - \frac{\hbar^2}{2M} \nabla^2 (\delta \psi_{\rm R} )
+ \frac{2 B}{\rho} (\delta \psi_{\rm R})
+ V \psi_0
\; \; \; .
\end{equation}

\noindent
The corresponding classical Hamiltonian is

\begin{displaymath}
{\cal H} = \int \biggl\{  V({\bf r}, t) ( \psi_0^2 + 2 \psi_0 
\delta \psi_{\rm R} )
+ \frac{2 B}{\rho} (\delta \psi_{\rm R} )^2
\end{displaymath}

\begin{equation}
+ \frac{\hbar^2}{2M} (
| {\bf \nabla} (\delta \psi_{\rm R} ) |^2 +
| {\bf \nabla} (\delta \psi_{\rm I} ) |^2 )
\biggr\} \; d{\bf r}
\; \; \; .
\end{equation}

\noindent
Fourier transforming the equations of motion we obtain

\begin{equation}
- i \hbar \omega \; \delta \hat{\psi}_{\rm R}
- \frac{\hbar^2 q^2}{2M} \; \delta \hat{\psi}_{\rm I} = 0
\end{equation}

\begin{equation}
(\frac{\hbar^2 q^2}{2M} + \frac{2B}{\rho} ) \; \delta \hat{\pi}_{\rm R}
- i \hbar \omega \; \delta \hat{\psi}_{\rm I} = \hat{V} \psi_0
\; \; \; .
\end{equation}

\noindent
For the static potential $V({\bf r}, t) = V_0 a^3
\delta^3 ({\bf r})$ we have $\hat{V} = 2 \pi V_0 a^3 \delta(\omega)$,
which gives $\delta \psi_{\rm I} = 0$ and

\begin{displaymath}
\delta \psi_{\rm R} ({\bf r} ) = -  \frac{ V_0 a^3 \; \psi_0}{(2\pi)^3}
\int \frac{1}{\hbar^2 q^2 / 2 M + 2 B / \rho } \; e^{i {\bf q}
\cdot {\bf r} } \; d{\bf r}
\end{displaymath}

\begin{equation}
= - V_0 a^3 \; \psi_0 \; \frac{2M}{\hbar^2} \frac{e^{- 2 r / \xi}}
{r} \; \; \; \; \; \; \;
(\xi = \frac{\hbar}{M v_s} )
\; \; \; .
\end{equation}

\noindent
The amount of fluid accumulated is thus

\begin{equation}
{\cal M} = M \int 2 \psi_0 \; \delta \psi_{\rm R} \; d{\bf r}
= - \frac{V_0 a^3 \rho}{v_s^2} \; \; \; .
\label{mass}
\end{equation}

\noindent
For the time-dependent potential

\begin{equation}
V({\bf r} , t) = V_0 a^3 \biggl\{
\delta^3 [{\bf r} - {\bf r}_0 (t) ]
+ \delta^3 [{\bf r} + {\bf r}_0 (t) ] \biggr\}
\; \; \; ,
\end{equation}

\noindent
where ${\bf r}_0 (t) = (\ell / 2) [ \cos(\omega_0 t) , \sin (\omega_0 t) ,
0 ]$ and $\omega_0 \ell << v_s$, only the quadrupole terms survive in the
far field, and we have

\begin{displaymath}
\delta \psi ( r ,  \theta , \phi ) \simeq \frac{V_0 a^3 \psi_0 \ell^2
\omega_0^3 \sin^2 (\theta )} {8 \pi \hbar v_s^4 r}
\end{displaymath} 

\begin{equation}
\times \biggl[ \frac{4 \hbar \omega_0}{Mv_s^2}
\cos(\chi) + i \sin(\chi) \biggr] \; \; \; ,
\end{equation}

\noindent
where $\chi = 2 \omega_0 (r / v_s - t)  - \phi$.  The radiated energy
flux is then

\begin{equation}
\frac{d{\cal P}}{d\Omega} \simeq \frac{(V_0 a^3 \psi_0 \ell^2
\omega_0^4)^2} {32 \pi^2 M v_s^9} \sin^4 (\theta) \; \; \; .
\end{equation}

\noindent
Combining this with Eq. (\ref{mass}) we obtain finally

\begin{equation}
{\cal P} \simeq \frac{({\cal M} \ell^2 \omega_0^3)^2}
{M \rho v_s^5} \frac{1}{16 \pi} \int_{-1}^1 (1 - \mu^2)^2 d\mu
\; \; \; ,
\end{equation}

\noindent
per Eq. (\ref{power}).

\end{document}